\documentstyle[11pt,AATS,epsf]{article}	

\begin{document}	

\title{Galaxy Formation:  One Star at a Time.  New Information 
from the Kinematics of Field Stars in the Galaxy 
} 

\author{Timothy C. Beers}
\affil{Dept. of Physics \& Astronomy, Michigan State University}
\author{Masashi Chiba}
\affil{National Astronomical Observatory of Japan}



\begin{abstract}
We present a review of recent studies of the formation and evolution of the
Milky Way galaxy, in particular based on large samples of non-kinematically
selected stars with available proper motions.  The Milky Way is argued to be a
reasonable template for the formation of large spiral galaxies, the only one in
which complete kinematical and abundance information can be readily obtained.
Ongoing and future projects to obtain proper motions and spectral information
for much larger samples of stars that will sharpen our perspective are
discussed. 
\end{abstract}


\section{Introduction}

Observational studies of galaxy formation often conjure visions of
deep optical and infrared images of luminous baryonic material caught ``in the
act'' at moderate to high redshift, perhaps inspired by the exquisite views
provided in the Hubble Deep Fields.  Although such data suggest that
at least the basic outlines of a hierarchical assembly model most likely can be
applied to many galaxies, the information content of these extremely faint
smudges on the night sky is at present limited by the resolution and
photon-gathering power of the largest telescopes.  By taking the position
that galaxy formation is best understood by detailed study of single galaxy,
rather than by obtaining only the sketchiest knowledge of the ancient history
of many galaxies, one is led to conclude that the optimal redshift for such
studies is at $z = 0$, i.e., the present.  We are speaking, of course, of our
home galaxy, the Milky Way.

In some respects, we can think of the Milky Way as an analog computer
simulation of the stages that many, if not most, large spiral galaxies must
have passed through, from the epoch of collapse from the general
expansion to the period of the first star formation, and the subsequent
evolution leading to the ``boundary conditions'' that the Milky Way presents
for modern observers.  There are numerous clues to the early history of galaxy
formation and evolution encoded in the motions and elemental abundances of
individual stars in the Galaxy.  The challenge to astronomers is to gather, and
most importantly, to interpret, this information to provide constraints on the
general patterns of (large) galaxy formation throughout the universe.  This
approach is not without its limitations, of course.  The Analog Milky Way V1.0
is a computer that:

\begin {itemize}
\item We live inside of
\item Can only directly show us a small portion of the ``program'' at once
\item We cannot re-boot
\item Runs its ``CPU'' at speeds of Gyrs, not GHz !
\end{itemize}

Nevertheless, the Milky Way is the only galaxy for which we can (relatively)
easily recover the full six-dimensional phase space information of position and
velocity for a substantial number of its luminous constituents while
simultaneously obtaining elemental abundance data for the same objects.
Moreover, our ability to gather such information is sure to increase rapidly in
the near future, with the completion of a number of full (or substantial) sky
coverage photometry programs (e.g., 2MASS, SDSS), the continuation and
extension of several ground-based astrometry programs (e.g., SPM, NPM, USNO),
the successful operation of the next generation of astrometric satellites
(e.g., FAME, SIM, GAIA), as well as the use of wide-field spectroscopic surveys
for the inspection of large numbers of spectra of stars in the field of the
Milky Way (e.g., 2DF and 6DF).

This conference was convened to discuss the broad issues of timescales in
astrophysics.  In the context of galaxy formation, this might be interpreted as
the estimation of the ages of the oldest (and perhaps more importantly, the
youngest) objects that might reasonably be assigned to the various luminous
components of the Milky Way that we presently recognize -- the halo, the
metal-weak thick disk (MWTD), the bulge, the thick disk, and the thin disk.  We
suggest that the first step toward such a goal is to at least identify the
proper {\it order of} formation of these components.  In this brief review, we
discuss how the kinematics and abundances of individual stars in a (still small)
sample of well-chosen stars can be used to make a tentative inference.  Many of
the questions we seek to answer will only completely yield when much larger
samples of stars with full information come available, but hopefully some of
the techniques we explore will provide a good starting point.

\section{Kinematic Studies -- Past and Present}

Kinematic studies of stellar populations in the Galaxy have long been limited
by the availability of large samples of stars with measurements of:

\begin {itemize}
\item Velocities (radial and tangential)
\item Distances  (in particular, consistent determinations)
\item Metallicities (accurate and consistent)
\end {itemize}

Such a database is required to constrain plausible scenarios for formation and 
evolution of the Milky Way.  A number of the issues under current discussion
include: 

\begin{itemize}

\item Measures of the local halo velocity ellipsoid and changes with
Galactocentric distance

\item The rotational character of thick disk and halo, and the existence of 
gradients in rotation velocity as a function of distance from the Galactic
plane

\item The existence of the MWTD, and the lower limit on its
metallicity 

\item Correlations (or lack thereof) between orbital eccentricity and the
metallicity of halo stars

\item The global density structure of the halo population, and estimates of the
axial ratios of the Galactic halo as a function of distance

\item Searches for ``kinematic substructure'' in the halo

\item Direct comparisons with simulations of galaxy formation

\end{itemize}

An ideal sample of stars for exploring these ideas would be unbiased in their
selection with respect to both kinematic properties and abundances throughout
the Galaxy.  Although a number of projects have been initiated that will
provide this sort of data (e.g., Morrison et al. 2000; Wyse et al. 2000), even
an approximation of the ideal has not yet been achieved.  Hence we must make a
choice between introduction of:

\begin {itemize}
\item Kinematic bias:  A tracer sample selected on the basis of stellar motions
in the Solar neighborhood, e.g., the proper motion selected samples of Ryan \&
Norris (1991) or Carney et al. (1994)

\item Abundance bias:  A tracer sample selected to include stars with
metallicities covering the range existing in the Galaxy ($-4.0 \le {\rm [Fe/H]}
\le 0.3$)

\end{itemize}

We have chosen to place our emphasis on the latter, on the grounds that
studies of the early formation history of the Galaxy must necessarily include
stars of extremely low metallicity, even though they represent a relatively
small fraction of the still-shining stars.  In this sense an abundance bias is
{\it required} in order to provide sufficient numbers of stars for meaningful
investigation, in particular of the halo population.

The study of large {\it non-kinematically selected} stellar samples 
began in earnest with the literature assemblage of Norris (1986), who studied
the kinematics of $\sim 800$ stars with available radial velocities, distances,
and abundances ${\rm [Fe/H]} \le -0.6$ (the upper cutoff being placed in order
to exclude ``spillover'' from the far more numerous thin disk stars).  Beers \&
Sommer-Larsen (1995) supplemented the Norris catalog with the inclusion of a
number of stellar samples with measured radial velocities, distances, and
abundances published in the literature in the intervening years (in particular
from the HK survey stars of Beers, Preston, \& Shectman 1992), finishing with a
sample of $\sim 1900$ stars.  The catalog of Beers et al. (2000) extended the
Beers \& Sommer-Larsen assemblage to include additional stars, in particular RR
Lyraes, and presented consistently determined estimates of distance and refined
radial velocities and abundances for a total sample in excess of $\sim 2000$
stars.  Of greatest importance, however, was the addition of proper-motion
information for over 50\% of the cataloged stars, based on several new
astrometric programs, including HIPPARCOS (ESA 1997), NPM (Klemola, Hanson, \& Jones 1993),
SPM (Platais et al. 1998),  STARNET (R\"oser 1996), ACT (Urban, Corbin, \& Wycoff 1998).
Additional information is now available
from the recently published Tycho--II catalog (H\o g et al. 2000), but it is not
considered in the discussion that follows.  Figure 1 shows the spatial
distribution of the stars in the Beers et al. catalog.

\begin{figure}[t]
\epsscale{0.6}
\plotone{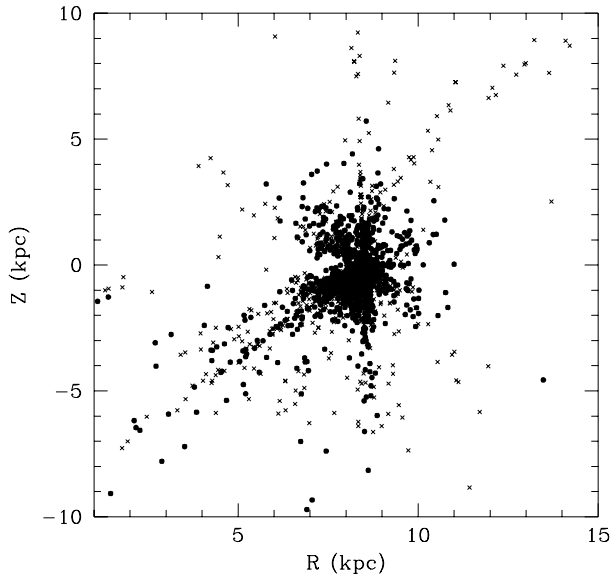}
\caption{Distribution of the sample of stars from Beers et al.
(2001) in the $(R,Z)$ plane.  The Sun is located at $R=8.5$ kpc, $Z = 0$ kpc.
Filled circles and crossed indicate the stars with and without available proper
motions, respectively.}
\end{figure}

\section{Results of the Chiba \& Beers Analysis}

Based on the large catalog of metal-poor stars with available proper motions,
Chiba \& Beers (2000) obtain full space motions for some 1200 stars.  The local
three-dimensional velocity components for this sample, $UVW$, are shown in
Figure 2 as a function of metallicity.  This is the first time that such
information has become available for an adequate number of low metallicity
stars chosen without kinematic bias.  

Inspection of panels (a) and (c) of Figure 2 suggests that there appears to be
present a ``core'' of stars with $-2.0 \le {\rm [Fe/H]}\le -0.6$ that is drawn
from a low-velocity-dispersion population, in addition to the usual
high-dispersion halo population.  This population is the MWTD,
a component of the Galaxy that was first suggested by Morrison, Flynn, \&
Freeman (1990), and was the subject of debate in the literature for some time.
Its presence now appears indisputable.

\begin{figure}[t]
\epsscale{0.5}
\plotone{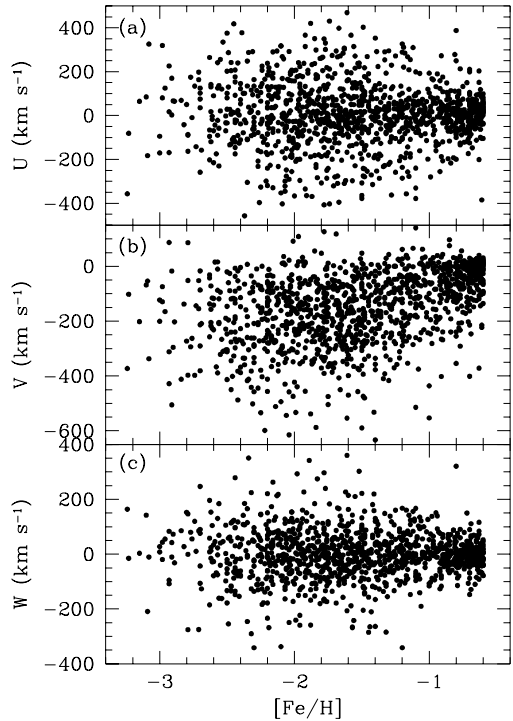}
\caption{Distribution of the velocity components $(U,V,W)$ vs.
[Fe/H] for the 1203 stars from Beers et al. (200) with available proper
motions.}
\end{figure}

There are other pieces of evidence that strongly suggest the presence of a MWTD
population.  For example, Figure 3 is a comparison of the cumulative
distribution of derived eccentricities for stars chosen in two abundance
regimes,  (a) for [Fe/H] $\le-2.2$, and (b) for $-1.4<$ [Fe/H] $\le-1$.  The
different lines correspond to the cases when the range of $|Z|$ is changed.
Panel (a) shows that even at quite low abundance, roughly 20\% of the stars
have $e < 0.4$.  The cumulative distribution of $e$ is unchanged when
considering subsets of the data with a range of $Z$, suggesting the absence of
any substantial disk-like component below [Fe/H] $= -2.2$.  By way of contrast,
panel (b) shows that stars with intermediate abundances exhibit (a) a higher
fraction of orbits with $e<0.4$ than for the lower abundance stars, (b) a
decrease in the relative fraction of low eccentricity stars as larger heights
above the Galactic plane are considered, and (c) convergence at larger heights
to a fraction that is close to the 20\% obtained for the lower abundance
stars.  These results imply that the orbital motions of the stars in the
intermediate abundance range are, in part, affected by the presence of a
MWTD component with a scale height on the order of 1 kpc.
\pagebreak

Chiba \& Beers obtain estimates of the local velocity ellipsoids of the halo
and thick disk components:
\begin {itemize}
\item halo:  $(\sigma_U,\sigma_V,\sigma_W)= (141\pm11,106\pm9,94\pm8)$
km~s$^{-1}$; [Fe/H] $< -2.2$
\item thick disk:  $(\sigma_U,\sigma_V,\sigma_W)=(46\pm4,50\pm4,35\pm3)$
km~s$^{-1}$; $-0.7 \le {\rm [Fe/H]}$
\end{itemize}

Progressing from higher to lower abundances, the velocity dispersions gradually
increase as one transitions from disk-like to halo-like behavior.  Under the
assumption that the MWTD population shares a common velocity ellipsoid with the
higher abundance thick-disk stars, a mixture model analysis suggests that, in
the Solar neighborhood, the MWTD contributes about 30\% of the metal-poor stars
in the abundance range $-1.7<$ [Fe/H] $\le-1$, and perhaps only 10\% below
[Fe/H] $= -1.7$.  It should be recalled, however, that the sample of stars
selected by Beers et al. (2000) were contributed primarily from
objective-prism surveys that generally placed a lower cut on Galactic
latitude of $|b| > 30\deg$.  Hence, the above fractional contributions of the
MWTD at intermediate abundances should be viewed as {\it lower limits} on the
actual fraction.  Indeed, a recent analysis of nearby metal-weak giants
selected with $|b| < 30\deg$ suggests that the fraction of MWTD stars at
abundances [Fe/H] $\le -1.7$ may be closer to $\sim 40\%$ (Beers et al. 2001),
much higher than the previously inferred value.  Clearly, data
obtained from surveys that also target lower Galactic latitudes are needed to
resolve this quandry (e.g., Wyse et al. 2000).

\begin{figure}[h]
\epsscale{0.5}
\plotone{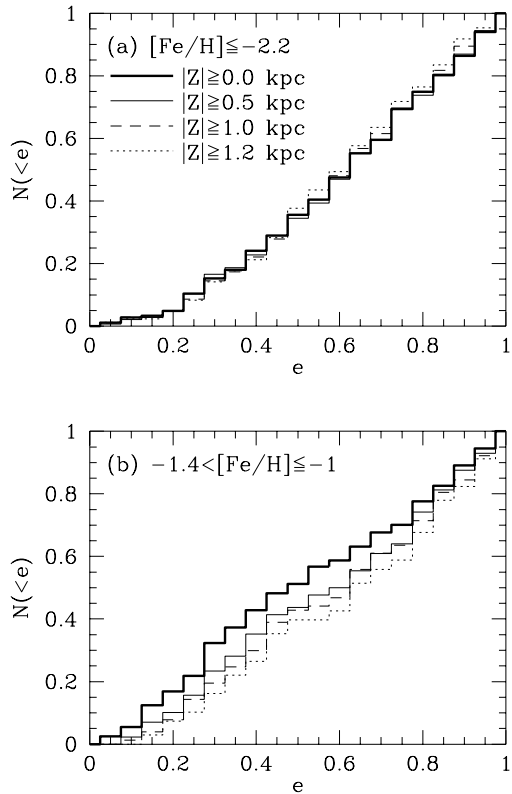}
\caption{Cumulative $e$ distributions, $N(<e)$, in the two abundance ranges (a) [Fe/H]
$\le-2.2$, and (b) $-1.4<$[Fe/H] $\le-1$. The thick solid, thin solid, dashed,
and dotted histograms denote the stars at $|Z|\ge0.0$ kpc (all stars),
$|Z|\ge0.5$ kpc, $|Z|\ge1.0$ kpc, and $|Z|\ge1.2$ kpc, respectively.}
\end{figure}

Figure 4 is a plot of the mean rotational velocity $<V_{\phi}>$ as a function of
[Fe/H] for stars selected from several cuts in distance above or below the
Galactic plane.  The general decline in rotation speed with decreasing [Fe/H]
marks the transition from a disk population (at the high abundance limit, the
canonical thick disk, at the low end, the MWTD) to a halo population.  A
``break'' at [Fe/H] $ \sim -1.7$ is evident in all three cuts in Z.  However,
note that below this metallicity, the rotational character of the cuts
seem to differ. This feature arises because, close to the Galactic plane, the
halo population exhibits a rather strong vertical rotational velocity gradient
$\Delta<V_\phi>/\Delta|Z|=-52 \pm 6$ km~s$^{-1}$ kpc$^{-1}$.  A smaller, but
still significant, gradient appears present in the thick-disk stars as well,
$\Delta<V_\phi>/\Delta|Z|=-30 \pm 3$ km~s$^{-1}$ kpc$^{-1}$).  The presence of
such gradients is a signature of dissipational collapse in both populations.

\begin{figure}[t]
\epsscale{0.7}
\plotone{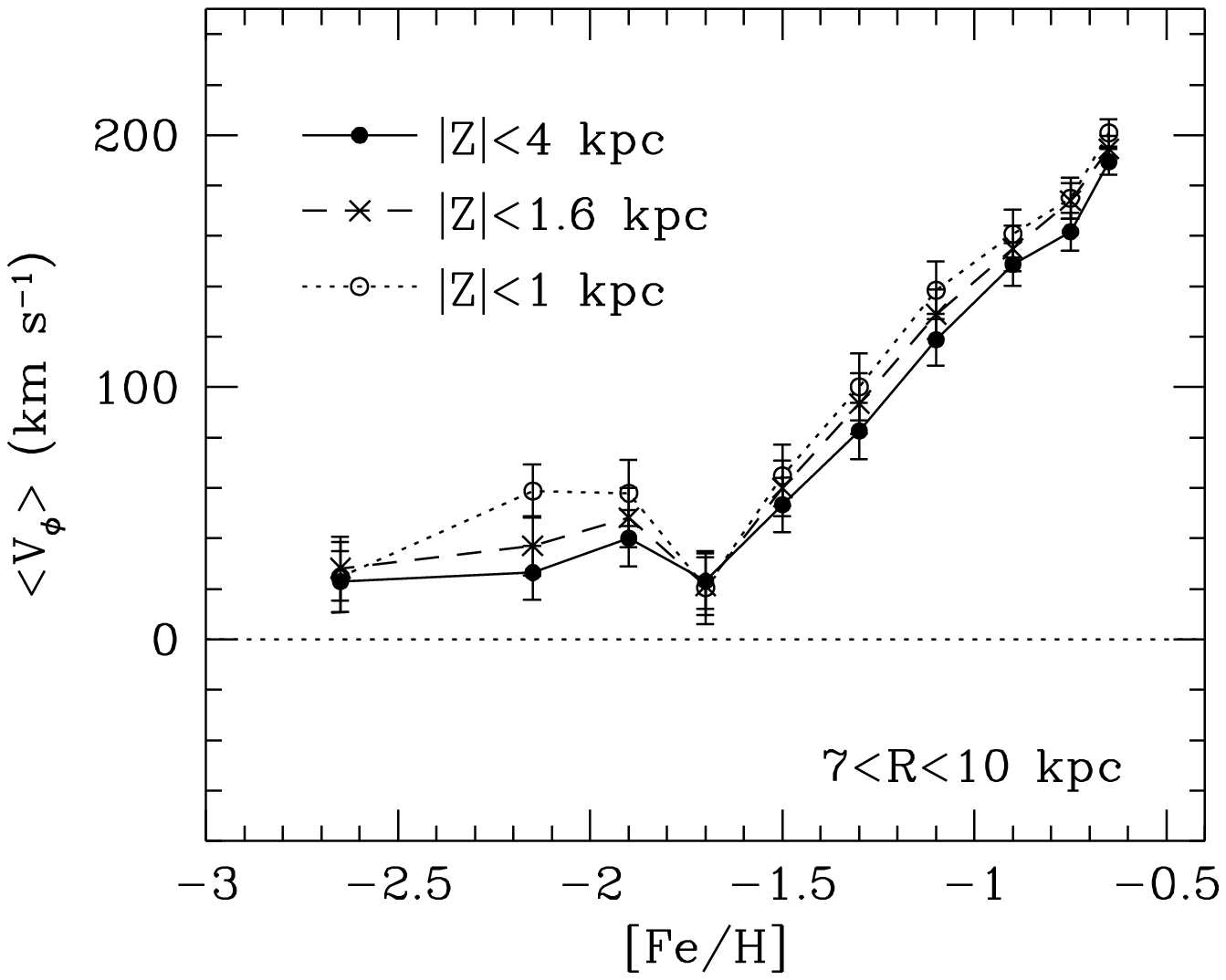}
\caption{Distribution of the mean rotational velocity
$<V_\phi>$ vs. [Fe/H] for stars closer than 4 kpc from the Sun, assuming an
$LSR$ rotation velocity of 220 km~s$^{-1}$. Filled circles, crosses, and open
circles correspond to the stars at $|Z|<4$ kpc, $|Z|<1.6$ kpc, and $|Z|<1$ kpc,
respectively. }
\end{figure}

Before closing this brief summary, it is interesting to view the entire
distribution of orbital eccentricities for the stars analysed by Chiba \& Beers
(2000).  Inspection of Figure 5 leaves little doubt that there exist metal-poor
stars with eccentricities that populate the entirety of the diagram.  This
result stands in rather stark contrast to previous claims of an existence of a
strong correlation between orbital eccentricity and metallicity (dating back to
the classic paper of Eggen, Lynden-Bell, \& Sandage 1962) that based their
initial selection of stars on high proper-motion surveys.

\begin{figure}[t]
\epsscale{0.8}
\plotone{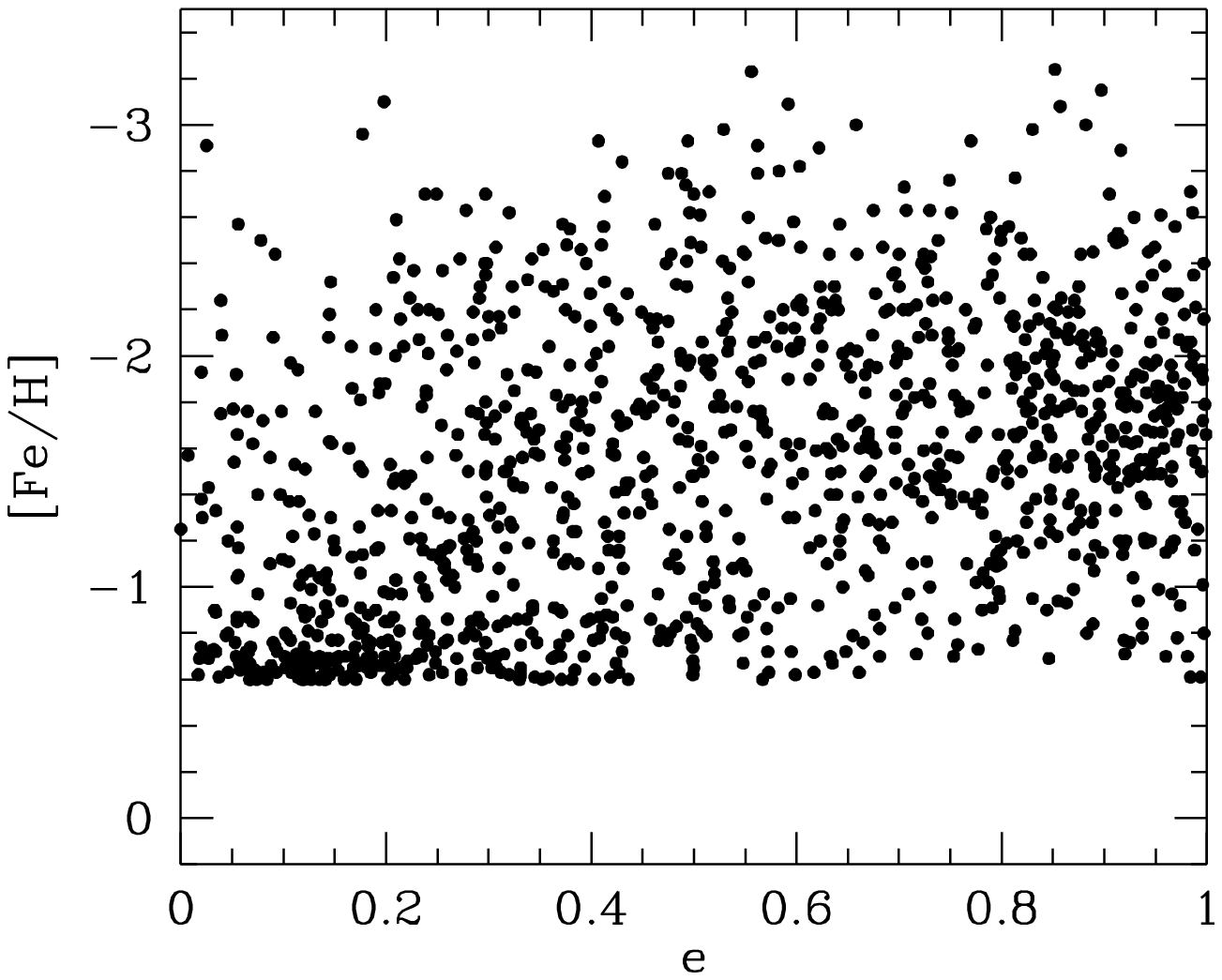}
\caption{The relation between [Fe/H] and $e$ for 1203 {\it non-kinematically
selected} stars with [Fe/H] $\le-0.6$.  Note the diverse range of $e$ even at
low metallicities. }
\end{figure}

\section{Reconstruction of the Halo Density Profile from Local Kinematics}

A number of authors have followed up the pioneering work of May \& Binney
(1986), who pointed out that an application of Jeans' theorem might enable the
reconstruction of the global structure of the stellar halo from kinematic
information of a reasonably large sample of local halo stars.  This technique
relies on the fact that halo stars that are presently found within a few kpc of
the Sun have, during their past motions, explored a substantial fraction of the
complete phase space of position and velocity in the Galaxy.  Sommer-Larsen \&
Zhen (1990) developed a maximum-likelihood methodology for the implementation of
this idea, and applied the technique to a sample of 118 nearby halo stars with
[Fe/H] $\le -1.5$.  Most recently, Chiba \& Beers (2000) have applied the
method to a sample of 359 stars with [Fe/H] $\le -1.8$ located within 4 kpc of
the Sun.  

Figure 6 shows the reconstructed density profile of the halo obtained by Chiba
\& Beers, compared to that derived by Sommer-Larsen \& Zhen, plotted as a
function of distance along the Galactic plane, $R$.  The density distribution
for $R > 8$ kpc is well described by a power-law profile $\rho \propto R^{-3.55
\pm 0.13}$, similar to that obtained by Sommer-Larsen \& Zhen except in the
outermost region, where the more recent analysis does not show a fall-off
(likely due to the smaller number of stars considered previously).  This result
is quite similar to previous estimates of the halo density profile based on
counts of globular clusters (Harris 1976; Zinn 1985), and field
horizontal-branch and RR Lyrae stars (Preston, Shectman, \& Beers 1991, and
references therein).  Note that the fall-off in reconstructed density for $R <
8$ kpc is an artifact arising from the lack stars in the Solar neighborhood
that explore the inner regions of the Galaxy.  

\begin{figure}[t]
\plotone{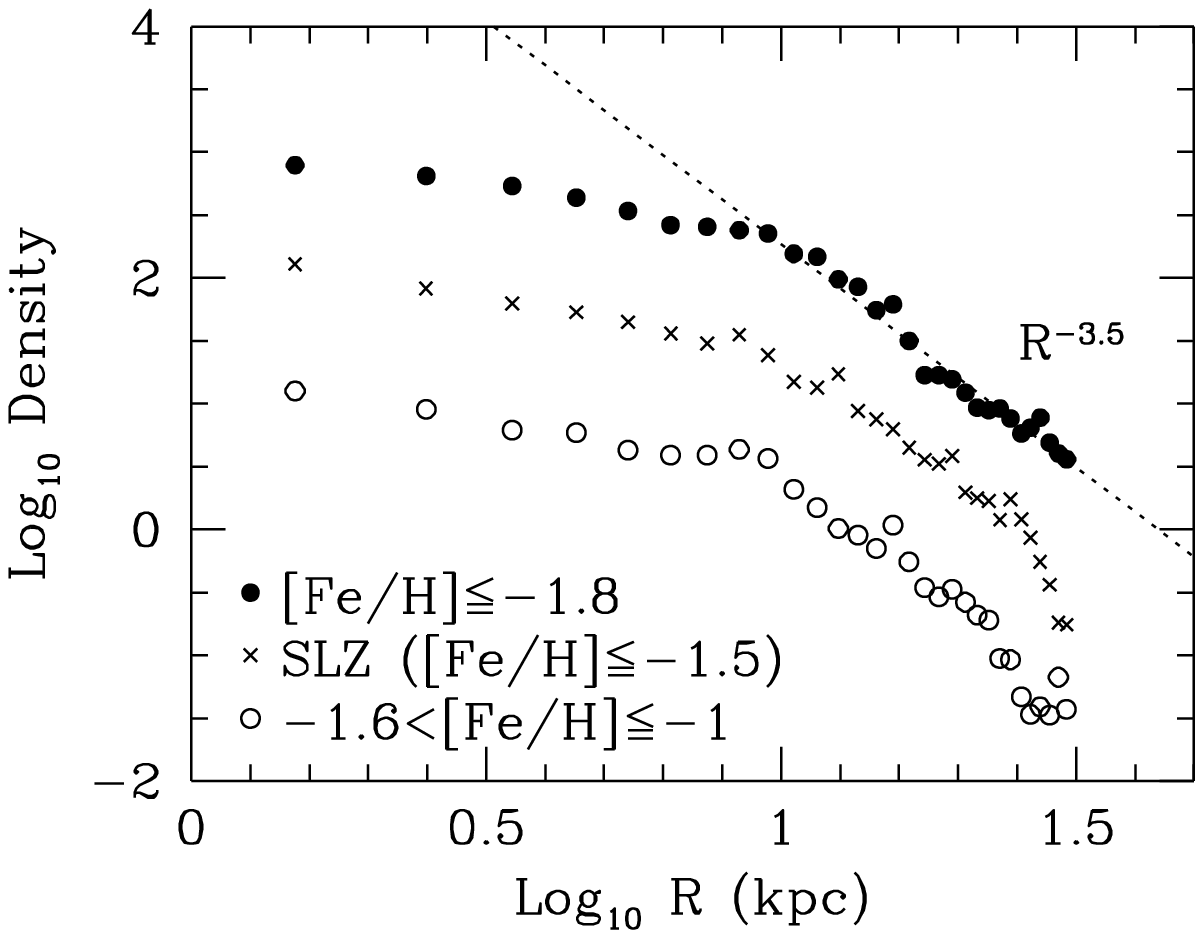}
\caption{Density distributions of the reconstructed halo in
the Galactic plane, for [Fe/H] $\le-1.8$ (filled circles) and $-1.6<$ [Fe/H]
$\le-1$ (open circles).  Both plots have been shifted arbitrarily along the
vertical axis.  The dotted line denotes the power-law model with exponent
$\beta=-3.5$.  For comparison, the density distribution based on the
Sommer-Larsen \& Zhen (1990) sample with [Fe/H] $\le-1.5$ is shown as crosses.}
\end{figure}

A different view of the reconstructed halo is presented in Figure 7a, where
contours of constant density are plotted in the $(R,Z)$ plane.  The appearance
of this figure is quite suggestive.  For the outer part of the halo, $R > 15$
kpc, the contours appear roughly spherical, while those in the inner halo
become increasingly flatter with declining distance.  A more quantitative
estimate of this behavior, based on the derived axial ratio of the
reconstructed halo as a function of distance, reveals that the density
distribution in the outer part of the halo, $R \sim 20$ kpc, is quite round.
The axial ratio $q$ appears to decrease with decreasing $R$ over $15<R<20$ kpc,
and the inner part, at $R<15$ kpc, exhibits $q \sim 0.65$.  Similar results,
obtained by completely different methods, were found by Preston et al. (1991)
and Morrison et al. (2000).  Thus, the present stellar halo can be described as
nearly spherical in the outer part and highly flattened in the inner part.

Chiba \& Beers (2001) extended this type of analysis to investigate the
structure of the Galactic halo at a time {\it before} the disk had a chance to
form.  Since the bulk of halo stars are found in the inner portion (owing to
the large exponent in the density profile), where the gravitational potential
arising from a disk dominates over that of the halo, the present-day halo might
have had its observed structure affected to a large degree by later disk
formation.  Chiba \& Beers studied the changes in the derived orbits of their
sample of stars when the potential associated with a disk is slowly removed
from their model, then reconstructed the density profile of the halo population
after it was completely absent.  Their result is shown in Figure 7b. 
The density contours in the (past) inner halo are substantially rounder then
inferred for the present halo.  The axial ratio obtained for the pre-disk halo
is approximately $q = 0.8$ for $R \le 15$ kpc, transitioning to spherical
for $R \sim 20$ kpc.  These results suggest that the presently flattened inner
halo is a consequence of both an initially slightly flattened inner
distribution that has been further flattened by later formation of a disk
structure.  An initially flattened halo is a signpost of early dissipative
formation of this component of the Galaxy.

\begin{figure}[t]
\epsscale{1.0}
\plotone{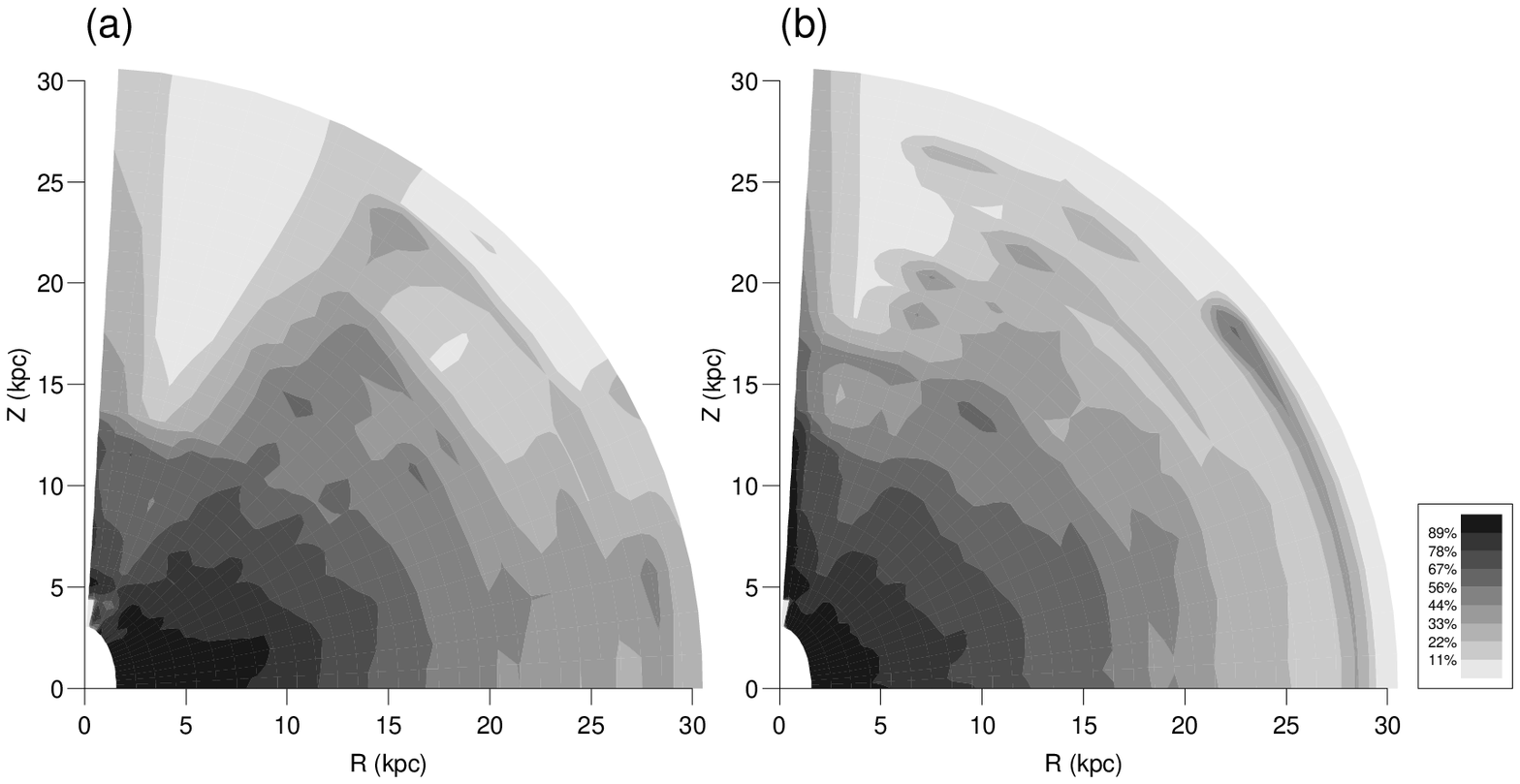}
\caption{Equidensity contours for the reconstructed halo in
the $(R,Z)$ plane, for stars with [Fe/H]$\le-1.8$.
Left and right panels correspond to the density distributions after and
before disk formation, respectively.}
\end{figure}

\section {Evidence for Halo Substructure}

Several studies (e.g., Helmi et al. 1999; Chiba \& Beers 2000) have presented
evidence for the existence of kinematic ``clumping'' of halo stars, a result
that is expected if a hierarchical assembly of the Galactic halo applies.
Though the numbers of stars in the recognized or suggested structures is
presently quite small, we anticipate that future (much) larger data samples
will demonstrate further evidence of the phenomenon.  Indeed, one of the first
exciting results from the SDSS reveals extensive spatially coherent structures
of field horizontal-branch and halo blue straggler stars, apparently associated
with the trail(s) of debris left by the Sagittarius dwarf (Yanny et al. 2000;
Ibata et al. 2001).  It should be noted that searches in angular momentum space
(obtainable for stars with full space motions) are inherently much more
sensitive than searches based on spatial association (Helmi \& White 1999), so
efforts to expand the numbers of stars with measured proper motions are of
particular value.  Studies of the elemental compositions of stars associated
with the identified kinematic structures (Ivans et al. 2001) will help to
establish if a ``chemical signature'' (such as low alpha-elements) can be
assigned to the process.

\section{Insights from Numerical Simulations}

The hypothesis of the dissipative formation of the inner flattened halo, as
well as the later accretion of satellites onto the outer halo, is a natural
consequence of the cold dark matter hierarchical clustering model (Bekki \&
Chiba 2000; 2001).  This model postulates that a protogalactic system initially
contains numerous subgalactic clumps, comprised of a mixture of primordial
abundance gas and dark matter.  Once star formation initiates in these clumps,
the energy associated with Type II SN explosions quickly drives the gas out
into the general ISM of a still-forming galaxy, at the same time touching off
formation of ``second- generation'' metal-poor stars in the dense shells of
these supernovae, as in the models of Tsujimoto, Shigeyama, \& Yoshii (1999).
The low-mass stars formed in this way are the long-lived fossils we now can
observe at extremely low metallicity.  Mergers of these small clumps, now
composed of dark matter and second-generation stars (but no gas), takes place
in a dissipationless manner.

Once the (enriched) gas that was expelled by early supernovae cools, it
will fall back into the ever-more massive clumps.  In the simulations of Bekki
\& Chiba, these larger clumps move gradually toward the central region of the
system, due to both dynamical friction and dissipative merging with smaller
clumps.  Finally, the last merging event occurs between the two most massive
clumps, and the metal-poor stars formed inside the clumps are disrupted and
spread over the inner part of the halo.  The aftermath is characterized by a
flattened density distribution.  Some fraction of the disrupted gas from the
clumps may settle into the central region of the system, and produce a more
enriched, more flattened density distribution, providing the ``raw material''
for the formation of the Galactic bulge and thick disk.  Some of the initially
small density fluctuations in the outer region would have gained systematically
higher angular momentum from their surroundings, and then slowly fall into the
system after most parts of the Galaxy were formed.  This may correspond to the
process of late satellite accretion, contributing primarily to the outer part
of the halo.  Thus, the reported initial state of the stellar halo can be
explained, at least qualitatively, in the context of a hierarchical clustering
scenario.

\section{Order of Formation of the Observed Luminous Components of the Galaxy}

The picture that emerges from the observations and simulations is one in which
the oldest stars that are presently observable in the Milky Way might be found
in any of several recognized components.  In this interpretation, the inferred
order of the formation of the components of the Milky Way is expected to be:

\begin{itemize}
\item Inner halo
\item Metal-weak thick-disk
\item Thick disk and Galactic bulge
\item Thin disk
\end {itemize}

All the while, the presently observed outer halo is in the process of
formation.  

We conclude that searches for the most metal-poor (and ancient) stars to be
found in the Galaxy should be concentrated on the inner halo.  This component
is expected to harbor the largest {\it relative} fraction of second-generation
stars.  However, it should be noted that the correlation between metal
abundance and age is expected to be extremely weak, even for the
most metal-deficient stars.  The enrichment history of the early Milky Way is
driven primarily by the efficiency of star formation during the complex early
epochs.  

In order to test the ordering suggested above, it is more revealing to consider
the {\it youngest} stars that might be associated with a given observed
component of the Galaxy, thus obtaining information on when star formation
ceased in that component.  Because the second-generation stars share a common
origin (in the shells of supernovae exploding in the first clumps of dark
matter and primordial gas), and then are re-distributed by subsequent
collisions of clumps, representatives of such stars might be found in all but
perhaps the thin disk of the Galaxy.

\section{The Importance of Future Spectroscopic and Astrometric Surveys}

There are several factors that limit to our ability to read back the history of
the formation of the Milky Way based on present data.  The ``resolution'' of
our vision is severely limited by the (still) relatively small numbers of stars
that are presently known with [Fe/H] $\le -2.0$.  The $\sim 1000$ stars below
this abundance limit that are currently recognized may {\it appear} to be a
sufficiently large number, but not when one is attempting, as we have, to
reconstruct the density structure of the early Galaxy from the phase space of
position and velocity that these stars explore.  Surveys that {\it efficiently}
identify much larger numbers of extremely metal-poor stars have recently been
completed, but are only now starting to be exploited to reveal the riches they
contain (e.g., the stellar component on the Hamburg/ESO survey, see Christlieb
\& Beers 2000 for a comparison of this survey with the HK-I survey, and the HK-II
survey, Rhee, Beers, \& Irwin 1999; Rhee 2000).  It should also be recognized
that stars of intermediate abundance, $-2.0 \le {\rm [Fe/H] \le -1.0}$ are
crucial for study of the MWTD-halo interface, and for that reason, are
deserving of attention as well.  Progress relies on the assignment of telescope
time to medium-resolution follow-up surveys, either with rapid single-star
measurements with 4m-class telescopes or with multi-fiber, wide-field
measurements with instruments such as the 6DF on the UK-Schmidt telesope
(Watson et al. 2000).

Even after these stars are recognized, and have had their radial velocities and
metallicities measured, there still remains the need for determination of
accurate proper motions and (eventually, from astrometric satellites to be
launched in the coming decade) accurate distances.  Only then will the complete
vision be realized.

What can be done in the meantime?  In recent years, until funding cuts
terminated progress, the most accurate ground-based proper-motion measurements
of the stars that occupy the MWTD and halo of the Galaxy have come from the
SPM, the Southern Proper Motion survey (e.g., Platais et al. 1998).  A proposal
to renew the SPM is currently under review at the (US) National Science
Foundation.  If this proposal is funded, the path is clear for the
determination of proper motions, as well as accurately calibrated $V$
magnitudes and $B-V$ colors, for the $\sim 10,000$ metal-poor (and $\sim
30,000$ field horizontal-branch and halo blue-straggler) stars in the southern
sky identified by the HK-I, HK-II, and Hamburg/ESO surveys, at least several
years before the first data return from future astrometric missions.  With this
data in hand, astronomers will be able to refine the issues they might hope to
explore with the wealth of data from the astrometric satellites.

We can, and hopefully will, be able to understand the complex process of galaxy
formation and evolution for at least ONE galaxy, our home, the Milky Way.
Unless the universe is perverse, what we learn in this process will apply to
many, if not most, large spiral galaxies.

\acknowledgments
TCB would like to thank the organizers of this meeting for partial financial
support that made his attendance possible, and the National Astronomical
Observatory of Japan for granting permission for his absence.  MC acknowledges
partial support from Grants-in-Aid for Scientific Research (09640328)
from the Ministry of Education, Science, Sports and Culture of Japan.


\begin{references}
\reference Beers, T.C., \& Sommer-Larsen, J. 1995, \apjs, 96, 175
\reference Beers, T.C., Preston, G.W., \& Shectman, S.A. 1992, \aj, 103, 1987
\reference Beers, T.C., Chiba, M., Yoshii, Y., Platais, I., Hanson, R.B., Fuchs, B., \& Rossi, S. 2000, \aj, 119, 2866
\reference Beers, T.C., Drilling, J.S., Rossi, S., Chiba, M., Rhee, J., \& F\"uhrmeister, B. 2001, in preparation.
\reference Bekki, K. \& Chiba, M. 2000, \apj, 534, L89
\reference Bekki, K. \& Chiba, M. 2001, \apj, submitted
\reference Carney, B.W., Latham, D.W., Laird, J.B., \& Aguilar, L.A. 1994, \aj, 102, 2240
\reference Chiba, M., \& Beers, T.C. 2000, \aj, 119, 2843
\reference Chiba, M., \& Beers, T.C. 2001, \apj, 549, 325 
\reference Christlieb, N., \& Beers, T.C. 2000, in HDS Workshop, eds. M.
Takada-Hidai \& H. Ando (Mitaka, Japan: NAO), p. 255 (astro-ph/0001378)
\reference Eggen, O.J., Lynden-Bell, D., \& Sandage, A.R. 1962, \apj, 136, 748
\reference ESA 1997, The Hipparcos and Tycho Catalogues (ESA SP-1200) (Noordwijk: ESA)
\reference Harris, W.E. 1976, \aj, 81, 1095
\reference Helmi, A., \& White, S.D.M. 1999, \mnras, 307, 495
\reference Helmi, A., Zhao, H.S., \& deZeeuw, P.T. 1999, in The Third Stromlo Symposium: The Galactic Halo, eds. B.K. Gibson, T.S. Axelrod, \& M.E.Putman (San Francisco:  ASP), 165, p. 125
\reference Helmi, A., White, S.D.M., deZeeuw, P.T., \& Zhao, H.S. 1999, Nature, 402, 53
\reference H\o g, E., Fabricius, C., Makarov, V. V., Urban, S., Corbin, T., Wycoff, G.,
Bastian, U., Schwekendiek, P., Wicenec, A. 2000, \aap, 355, L27
\reference Ibata, R., Irwin, M., Lewis, G.F., \& Stolte, A. 2001, \apj, 547, L133
\reference Ivans, I.I., Chiba, M., Beers, T.C., \& Sneden, C. 2001, in preparation
\reference Klemola, A.R., Hanson, R.B., \& Jones, B.F. 1993, Lick Northern Proper Motions Program:  NPM1 Catalog (NSSDC A1199), (Washington, D.C.:  NASA)
\reference May, A. \& Binney, J. 1986, \mnras, 221, 857
\reference Morrison, H.L., Flynn, C., \& Freeman, K.C. 1990, \aj, 100, 1191 
\reference Morrison, H.L., Mateo, M., Olszewski, E.W., Harding, P.,
Dohm-Palmer, R.C., Freeman, K.C., Norris, J.E., \& Morita, M. 2000, \aj, 119, 2254
\reference Norris, J.E. 1986, \apjs, 61, 667
\reference Platais, I., Girard, T.M., Kozhurina-Platais, V., Van Altena, W.F.,
Lopez, C.E., Mendez, R.A., Ma, W., Yang, T., MacGillivray, \& Yentis, D.J. 1998, \aj, 116, 2556
\reference Preston, G.W., Shectman, S.A., \& Beers, T.C. 1991, \apj, 375, 121
\reference Rhee, J. 2000, Ph.D. Thesis, Michigan State University
\reference Rhee, J., Beers, T.C., \& Irwin, M.J. 1999, BAAS, 194, 84.11 
\reference R\"oser, S. 1996, in IAU Symposium No. 172, eds. S. Ferraz--Mello et
al., (Dordrecht:  Kluwer), p. 481
\reference Ryan, S.G., \& Norris, J.E. 1991, \aj, 101, 1835
\reference Sommer-Larsen, J., \& Zhen, C. 1990, \mnras, 242, 10
\reference Tsujimoto, T., Shigeyama, Y., \& Yoshii, Y. 1999, \apj, 519, L63
\reference Urban, S.E., Corbin, T.E., \& Wycoff, G.L. 1998, \apj, 115, 2161
\reference Watson, F.G., Parker, Q.A., Bogatu, G., Farrell, T.J., Hingley,
B.E., \& Miziarski, S. 2000, Proc. SPIE, 4008, p. 123
\reference Wyse, R.F.G., Gilmore, G., Norris, J.E., \& Freeman, K.C. 2000,
BAAS, 197, 04.24
\reference Yanny, B. et al. 2000, \apj, 540, 825
\reference Zinn, R. 1985, \apj, 293, 424

\end{references}
\end{document}